\documentclass[epj]{svjour}
\usepackage{graphics}
\begin{document}
\title{Prolate over oblate dominance in deformed nuclei as a consequence of the SU(3) symmetry and 
the Pauli principle}
\author{Dennis Bonatsos
}                     % Do not remove
\institute{Institute of Nuclear and Particle Physics, National Centre for Scientific Research 
``Demokritos'', GR-15310 Aghia Paraskevi, Attiki, Greece}
\date{Received: date / Revised version: date}
\abstract{
We show that the dominance of prolate over oblate shapes in even-even deformed nuclei
can be derived from the SU(3) symmetry and the Pauli principle.  
 % end of PACS codes
} %end of abstract

\titlerunning{Prolate over oblate dominance in deformed nuclei}
\maketitle
%
%\section{Introduction}
%\label{intro}

As remarked in  a recent review \cite{HM} by Hamamoto and Mottelson, ``the observed almost complete dominance of prolate over oblate deformations in the ground states of deformed even-even nuclei is 
not yet adequately understood''. 

In this Letter we suggest that the appearance of prolate (rugby ball shaped) deformations in the
ground states of the majority of deformed even-even nuclei, as opposed to the existence of a few oblate (pancake shaped) ground state deformations 
appearing just below the simultaneous closing of the proton valence shell and the neutron valence shell can be understood in terms of the SU(3) symmetry and the Pauli principle alone. 

The basic steps needed in this proof are listed here.

1) The SU(3) symmetry discovered by Elliott \cite{Elliott1,Elliott2} in the sd shell, is destroyed 
in higher shells by the presence of the strong spin-orbit interaction. However, it has been recently 
realized \cite{Martinou} that the SU(3) symmetry can be approximately recovered in the higher shells
by substituting the opposite  parity orbitals (except the one lying highest in energy) invading a given shell from above by the normal parity orbitals which have deserted this given shell in order to take refuge in the shell below. Using the Nilsson asymptotic quantum numbers \cite{Nilsson1,Nilsson2} one can see that during this replacement all angular momentum projections (projections of orbital angular momentum, spin, total angular momentum) remain unchanged, while only the total number of quanta $N$ and the number of quanta in the $z$-direction, $n_z$, are reduced by one unit. This approximation is in line with the remark by Mottelson \cite{Mottelson} that the asymptotic quantum numbers of the Nilsson model can be seen as a generalization of Elliott's SU(3). 

2) The SU(3) symmetry being restored, it is known that the ground state will belong to the highest weight irreducible representation $(\lambda,\mu)$ of SU(3) \cite{Elliott1,Elliott2}, occurring in the reduction from the U(N) symmetry of the relevant (restored) harmonic oscillator shell to its SU(3) subalgebra. The sd shell is characterized by the U(6) overall symmetry, while the pf, sdg, pfh, and sdgi shells, which occur by the procedure described in 1) in the place of the 28-50, 50-82, 82-126, and 126-184 shells, correspond to the U(10), U(15), U(21), and U(28) symmetries, respectively, all of which possess SU(3) subalgebras \cite{BK}.

3) The determination of the highest weight irreducible representation (irrep), as well as of the other irreps occurring in a reduction from U(N) to SU(3) is a formidable mathematical task, which, however, has been solved,
with a code being available \cite{code}. Early tabulations for U(6) and U(15) can be found in Refs. 
\cite{Elliott1,Perez,Brahmam} respectively.

4) In the tabulations mentioned in 3), and/or using the code of Ref. \cite{code} one can easily see
that the particle-hole symmetry is strongly broken in the shells higher than the sd shell. In particular, the particle-hole symmetry is valid only up to 4 particles or 4 holes. This can also be seen in the partial results for U(10) reported in Table 5 of Ref. \cite{pseudo1}. The breaking of the particle-hole symmetry cannot be seen in even-even nuclei in the sd shell, because of its small size. 

5) The highest weight irrep for a given number of valence particles can also be determined by considering the correct order of the Nilsson orbitals for a given deformation and gradually filling them until all particles available are accommodated. One can then easily determine the quantity 
$2\lambda+\mu$ for the highest weight irrep by finding the maximum value of the sum $3\sum n_z -\sum N$ for the occupied Nilsson orbitals \cite{Elliott2,pseudo1}. The loss of particle-hole symmetry is already visible at this point. The determination of $\mu$ for the highest weight irrep requires the determination of the highest value of $\sum n_x -\sum n_y$ subject to the restriction of $\sum n_z$ used before \cite{pseudo1}. At this point, the use of the code of Ref. \cite{code} becomes necessary 
for non-trivial numbers of valence particles. 
  
6) Irreps with $\lambda > \mu$ correspond to prolate shapes, while irreps with $\lambda < \mu$
represent oblate shapes. One way to derive this result is to consider the Bohr collective model 
\cite{BM} and SU(3), and require a linear mapping between the invariant operators of the two models,
thus obtaining relations connecting the collective variables $\beta$ and $\gamma$ to the SU(3) quantum numbers $\lambda$ and $\mu$ \cite{Evans,Castanos,Park}.

7) If one considers the rare earths with 50-82 protons and 82-126 neutrons, from the number of the valence protons one can determine in U(15), according to 3), the relevant highest weight irrep for protons, $(\lambda_p, \mu_p)$. In the same way, one can determine in U(21) the relevant highest weight irrep for neutrons, $(\lambda_n, \mu_n)$. Then the highest weight irrep for this specific nucleus will be $(\lambda_p+\lambda_n, \mu_p+\mu_n)$. 

8) Working out the details one can see \cite{Sarantopoulou} that this procedure leads to a 
prolate-to-oblate shape/phase transition in the W, Os, and Pt isotopes at N=116, the first 
oblate isotopes in these series appearing at $N=116$, in agreement with empirical observations.
In particular, $^{190}$W \cite{Alkhomashi} and $^{192}$Os \cite{Namenson} have been suggested as lying at the prolate-oblate border, while for $^{194}$Os \cite{Wheldon} and $^{198}$Os \cite{Podolyak} a clearly oblate character has been found. The collection of data 
of the chain of even nuclei (differing by two protons or two neutrons) $^{180}$Hf, $^{182-186}$W, $^{188-192}$Os, $^{194,196}$Pt, $^{198,200}$Hg,
considered in Ref. \cite{Linnemann}, also suggests that the transition occurs between $^{192}$Os and $^{194}$Pt. Notice that all experimental 
information cited here is in agreement with the present theoretical predictions, since $^{190}$W, $^{192}$Os, and $^{194}$Pt are $N=116$ isotones. It is therefore seen that in the rare earth region 
the majority of deformed nuclei are prolate, with the exception of a few nuclei just below 
the $Z=82$ proton shell closure and the $N=126$ neutron shell closure. 
Similar conclusions are drawn  also for the rare earths with 50-82 protons and 50-82 neutrons,
with the prolate-to-oblate transition occurring at $N=72$ in the W, Os, and Pt isotopic chains \cite{Sarantopoulou}. 
 
Some further remarks are in place.

1) Recent self-consistent Skyrme Hartree-Fock plus BCS calculations \cite{Sarriguren} and Hartree-Fock-Bogoliubov calculations \cite{Robledo,Nomura83,Nomura84} studying the structural evolution in neutron-rich Yb, Hf, W, Os, and Pt isotopes also suggest the $N \approx 116$ nuclei in this region as lying at the transition points between prolate and oblate shapes, in agreement to the findings mentioned in 8).  
 
2)The approximate SU(3) symmetry described in 1) is complementary to the pseudo-SU(3) symmetry \cite{pseudo1,pseudo2,Ginocchio}, in which the normal parity orbitals are approximated, while the intruder parity orbitals remain intact, in contrast to the present approximate symmetry, in which the normal parity orbitals remain intact, while the intruder parity orbitals are approximated. 

In conclusion, the prolate-over-oblate dominance in the ground states of even-even deformed nuclei appears to be rooted in the SU(3) symmetry, approximately valid in the shells 
higher than the sd shell, and in the gradual filling by the valence protons and neutrons of the Nilsson orbitals according to the restrictions imposed by the Pauli principle.

\end{document}